\documentclass[aps,prl,preprint,showpacs,superscriptaddress,unsortedaddress,groupedaddress,altaffillsymbol]{revtex4}
\usepackage{braket}
\usepackage{upgreek}
\usepackage{graphicx}
\usepackage{amsmath}
\usepackage{bm}
\usepackage{amssymb} 

\usepackage[usenames,dvipsnames]{xcolor}
\usepackage{soul}

\newcommand{\tc}[1]{\protect\textcircled{\protect\raisebox{-0.7pt}{\small{#1}}}}

\begin{document}
\title{Error Compensation of Single-Qubit Gates in a Surface Electrode Ion Trap Using Composite Pulses}
\date{\today}

\author{Emily Mount}
\affiliation{Electrical and Computer Engineering Department, Duke University, Durham, NC 27708, USA}

\author{Chingiz Kabytayev}
\affiliation{Schools of Chemistry and Biochemistry, Computational Science and Engineering, and Physics, Georgia Institute of Technology, Atlanta, GA 30332, USA}

\author{Stephen Crain}
\affiliation{Electrical and Computer Engineering Department, Duke University, Durham, NC 27708, USA}

\author{Robin Harper}
\affiliation{School of Physics, University of Sydney, Sydney, New South Wales 2006, Australia}

\author{So-Young Baek}
\affiliation{Electrical and Computer Engineering Department, Duke University, Durham, NC 27708, USA}

\author{Geert Vrijsen}
\affiliation{Electrical and Computer Engineering Department, Duke University, Durham, NC 27708, USA}

\author{Steven T. Flammia}
\affiliation{School of Physics, University of Sydney, Sydney, New South Wales 2006, Australia}

\author{Kenneth R. Brown}
\affiliation{Schools of Chemistry and Biochemistry, Computational Science and Engineering, and Physics, Georgia Institute of Technology, Atlanta, GA 30332, USA}

\author{Peter Maunz}
\affiliation{Sandia National Laboratories, Albuquerque, NM 87123, USA}

\author{Jungsang Kim}
\affiliation{Electrical and Computer Engineering Department, Duke University, Durham, NC 27708, USA}

\begin{abstract}
The fidelity of laser-driven quantum logic operations on trapped ion qubits tend to be lower than microwave-driven logic operations due to the difficulty of stabilizing the driving fields at the ion location. Through stabilization of the driving optical fields and use of composite pulse sequences, we demonstrate high fidelity single-qubit gates for  the hyperfine qubit of a $^{171}\text{Yb}^+$ ion trapped in a microfabricated surface electrode ion trap. Gate error is characterized using a randomized benchmarking protocol, and an average error per randomized Clifford group gate of $3.6(3)\times10^{-4}$ is measured. We also report experimental realization of palindromic pulse sequences that scale efficiently in sequence length.

\end{abstract}

\maketitle

The trapped atomic ion qubits feature desirable properties for use in a quantum computer such as long coherence times~\cite{Langer2005}, high fidelity qubit measurement~\cite{Noek2013}, and universal logic gates~\cite{Home2009}. The quality of quantum logic gate operations on trapped ion qubits has been limited by the stability of the control fields at the ion location used to implement the gate operations. For this reason, the logic gates utilizing microwave fields~\cite{Brown2011, Shappert2013, Harty2014} have shown gate fidelities several orders of magnitude better than those using laser fields~\cite{Knill2008, Benhelm2008, Ballance2014}. The UV laser beams  used to drive Raman gates for a hyperfine ion qubit present a major challenge as they are subject to severe wavefront distortion in air due to turbulence, leading to amplitude and phase fluctuations at the ion location.

Microfabricated surface electrode ion traps, where atomic ions are trapped above a two dimensional surface of electrodes, can provide a scalable platform on which to build an ion-based quantum computer~\cite{Chiaverini2005,Kim2005a}. Experiments using surface traps have demonstrated coherence times of more than 1 second~\cite{Mount2013}, state detection with fidelities greater than 99.9\%~\cite{Noek2013}, and low error single-qubit gates ($\epsilon \lesssim 2.0(2)\times10^{-5}$) using integrated microwave waveguides~\cite{Brown2011, Harty2014}. Use of high power UV lasers close to the trap surface can lead to substantial charging due to unwanted exposure~\cite{Harlander2010}. The recent development of single-mode fibers capable of delivering high power UV laser beams~\cite{Colombe2014} opens the possibility of significantly reducing the free-space UV beam path length and delivering a clean spatial mode to the ions, eliminating unwanted scattering off nearby trap structures.

Here, we demonstrate low-error single-qubit gates performed using stimulated Raman transitions on an ion qubit trapped in a microfabricated chip trap. Gate errors are measured using a randomized benchmarking protocol~\cite{Knill2008, Wallman2014, Magesan2012a}, where amplitude error in the control beam is compensated using various pulse sequence techniques~\cite{Wimperis1994, Low2014}. Using B2 compensation~\cite{Wimperis1994}, we demonstrate single-qubit gates with an average error per randomized Clifford group gate of $3.6(3)\times10^{-4}$. We also show that compact palindromic pulse compensation sequences (PD\textit{n})~\cite{Low2014} compensate for amplitude errors as designed.

\begin{figure}
\begin{center}\includegraphics[width=\textwidth]{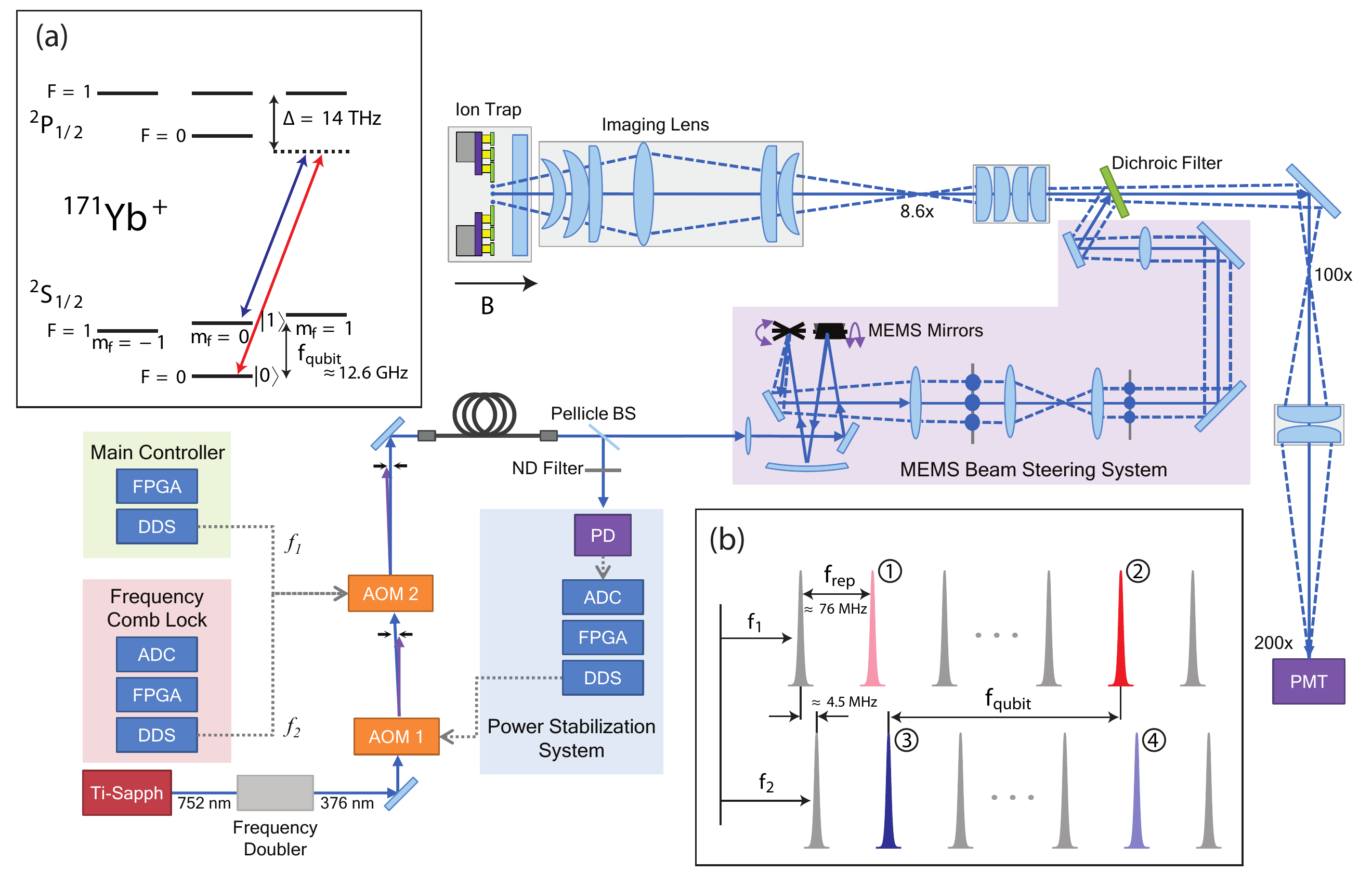}
\caption{
Experimental setup. Inset (a) shows the relevant energy levels used in the $^{171}$Yb$^+$ ion. Raman beams from a frequency-doubled picosecond Ti-Sapph laser passes through AOM 2 driven by two frequencies ($f_{1}$ and $f_{2}$), and creates the two optical frequency combs as shown in inset (b).  See text for a detailed description of the setup. DDS: Direct digital synthesizer, ADC: Analog-to-digital converter, PD: Photodiode.
}
\label{fig:fig1}
\end{center}
\end{figure}

Two hyperfine ground states of the $^{171}\text{Yb}^+$ ion ($\ket{0} \equiv\,^2$S$_{1/2} {\left|F = 0, m_f = 0\right>}$  and $\ket{1} \equiv\, ^2$S$_{1/2}{\left|F = 1, m_f = 0\right>}$, shown in Fig.~\ref{fig:fig1}a, separated by $f_{\text{qubit}}\approx 12.6$ GHz, serve as our qubit states. The energy separation between these states is relatively insensitive to the magnetic field fluctuations at the relevant magnetic field ($\sim 3$~kHz/Gauss at 5 Gauss). Continuous wave (CW) lasers are used to perform Doppler cooling, resonant scattering for qubit state detection, and optical pumping out of the $^2\text{D}_{3/2}$  state (not pictured in Fig.~\ref{fig:fig1}a). For Raman transitions, we use picosecond pulses from a mode-locked titanium-sapphire (Ti:Sapph) laser frequency-doubled to a center wavelength of 376~nm, which creates combs in the frequency domain with comb teeth spacing equal to the laser repetition rate ($f_{\text{rep}} \approx 76$~MHz).  The frequency doubler output is split into two nearly co-propagating frequency combs using a single acousto-optic modulator (AOM2) driven with modulation frequencies $f_1$ and $f_2$, as shown in the inset Fig.~\ref{fig:fig1}b.  Resonant transitions are driven by pairs of optical frequency comb teeth (\tc{2} and \tc{3} in Fig.~\ref{fig:fig1}b), one from each comb, with a frequency difference equal to $f_{\text{qubit}}$~\cite{Mount2013,Hayes2010}. As the repetition rate $f_{\text{rep}}$ of the laser drifts, the frequency difference between these two comb teeth is actively stabilized to match $f_{\text{qubit}}$ by adjusting the modulation frequency $f_2$ to maintain the quantity $166\times f_{\text{rep}}+f_{2}$ constant~\cite{ Mount2015,Islam2014}. After the AOMs, the Raman beams are delivered through a 3-m single-mode fiber to a micro-electromechanical systems (MEMS)-based laser beam steering system that is capable of fast ($< 2 \mathrm{\mu}$s), re-configurable addressing of individual ions in a linear chain with low crosstalk ($<3\times10^{-4}$)~\cite{Crain2014}. The Raman laser power delivered to the MEMS system is actively stabilized using a gated digital proportional-integral (PI) loop (see below)~\cite{Mount2015}. After the addressing system, the co-propagating linear polarization of the Raman beams is converted to circular polarization with a quarter-wave plate to drive a $\sigma^+$-transition.  A dichroic filter, which reflects 376 nm light and transmits 370 nm light, is used to fold the Raman beams into a high numerical aperture (NA~$=0.6$, PhotonGear) lens that serves the dual purpose of imaging the ion fluorescence and focusing the frequency combs onto the ion. For state detection, the ion is imaged at 200$\times$ magnification onto a photomultiplier tube (PMT) array. All experimental timing and measurement recording is carried out by a field-programmable gate array (FPGA) located in our main controller (shown in Fig.~\ref{fig:fig1}).

For each experiment, the ion is first Doppler cooled for 1~ms using light that is red-detuned from $^2\text{S}_{1/2}\ket{F\!=\!1}\!\rightarrow\!^2\text{P}_{1/2}\ket{F\!=\!0}$ resonance.  The qubit is then initialized to the $\ket{0}$ state by applying CW light resonant with $^2\text{S}_{1/2}\ket{F\!=\!1}\!\rightarrow\!^2\text{P}_{1/2}\ket{F\!=\!1}$ transition for 20~$\mathrm{\mu}$s. Following initialization, the intensity-stabilized Raman beams are pulsed on and off using the AOMs for a duration corresponding to the gate implementation. To measure the qubit state, light resonant with $^2\text{S}_{1/2}\ket{F\!=\!1}\!\rightarrow\!^2\text{P}_{1/2}\ket{F\!=\!0}$ transition is turned on for 400~$\mathrm{\mu}$s while ion fluorescence is measured using the relevant pixel of the PMT array. The ion will fluoresce if the qubit is in the $\ket{1}$ state, while the $\ket{0}$ state remains dark.

Raman beam intensity fluctuations at the ion location result in systematic amplitude errors which decrease gate fidelity. To minimize intensity fluctuations at the ion, a power stabilization system is implemented~\cite{Mount2015}, where a small amount of collimated light is picked-off after the beam exits the fiber and is incident on a photodiode (PD). The PD provides the error signal for a proportional-integral (PI) feedback loop that controls the amplitude of the direct digital synthesizer (DDS) signal driving AOM 1, as shown in Fig.~\ref{fig:fig1}. The PI loop is gated by a digital trigger pulse sent from the FPGA in our main controller. The Raman beams and PI loop are turned on during the Doppler cooling cycle, allowing the power to be stabilized. The loop is then turned off for the remainder of the experiment. Using this approach, the intensity fluctuations due to beam pointing instability before the fiber are corrected. Beam pointing instability after the fiber, due to air currents, is passively suppressed by enclosing the entire experiment in a box.

A composite pulse sequence can be used in place of a single Raman pulse to make it robust against systematic errors such as amplitude, timing, crosstalk, or detuning errors~\cite{Wimperis1994,Low2014,Merrill2011a,Soare2014,Merrill2014} and time-dependent control errors~\cite{Kabytayev2014}. In our experiment, the impact of residual systematic amplitude errors in the Raman beams is suppressed through the use of compensating pulse sequences. Since these techniques are usually designed to work on systematic errors that are  constant over the duration of the sequence, the sequence length determines the bandwidth below which the effect of fluctuating error is suppressed~cite{Kabytayev2014}. The length of most compensating pulse sequences increases rapidly at higher error correction order. The palindromic pulse sequences (PD{\it n}) are unique in that they scale linearly with the corrected error order (to $n=12$)~\cite{Low2014}. Here we analyze the use of B2~\cite{Wimperis1994} and PD6~\cite{Low2014} composite pulses and their ability to correct static amplitude errors in the presence of additional phase and timing errors for Clifford group gates.

In the absence of noise, the target rotation $R(\theta, \phi)$ rotates the Bloch vector by angle $\theta$ around the axis $\sigma_{\phi} \equiv X \cos \phi + Y \sin \phi$, represented by a unitary propagator $R(\theta, \phi) = \exp \left (-\frac{i}{2} \theta \sigma_{\phi} \right )$. The B2 compensation sequence (also known as BB1), introduced by Wimperis~\cite{Wimperis1994}, is designed to correct the errors in the pulse area (due to amplitude or timing errors) to $O(\epsilon^2)$, where $\epsilon$ is the fractional error in the control signal~\cite{Merrill2011a}. B2 compensation translates each single rotation into a sequence of 4 rotations. A target rotation $R(\theta_{t},\phi_{t})$ becomes
\begin{equation}\label{eq:bb1}
R(\theta_{t},\phi_{t}) \xrightarrow{B2} R(\theta_{t},\phi_{t})  R(\pi,\phi_{t}+\phi_{B2}) R(2\pi,\phi_{t}+3\phi_{B2}) R(\pi,\phi_{t}+\phi_{B2}),
\end{equation}
where 
\begin{equation}\label{eq:phibb1}
\phi_{B2}=\text{cos}^{-1}\left(\frac{-\theta_{t}}{4\pi}\right).
\end{equation}
A single B2 compensated pulse requires a  total rotation angle of $\theta_{total}=4\pi+\theta_{t}$, which requires time $t_{total} = \frac{\theta_{total}}{\Omega}$, where $\Omega$ is the Rabi frequency.  While the B2 sequence is fairly short at $n=2$, the B{\it n} sequences increase in length exponentially with increasing {\it n}, requiring $O(\exp(n^2))$ pulses to reduce amplitude errors to $O(\epsilon^n)$~\cite{Brown2004}.

In comparison to B{\it n} sequences, palindromic compensation sequences (PD{\it n}) scale efficiently in length, requiring only $2n$ pulses ($\theta_{total}=2n\pi+\theta_{t}$) to cancel errors to $n^{\text{th}}$ order, up to $n=12$~\cite{Low2014}.  Here we use PD6 ($n=6$), where a target rotation $R(\theta_{t},\phi_{t})$ is replaced by
\begin{multline}\label{eq:pd6}
R(\theta_{t},\phi_{t}) \xrightarrow{PD6} R(\theta_{t},\phi_{t})  R(\pi,\phi_{t}+\phi_{PD6:1})   R(\pi,\phi_{t}+\phi_{PD6:2})   R(\pi,\phi_{t}+\phi_{PD6:3})   R(\pi,\phi_{t}+\phi_{PD6:4})\\
  R(\pi,\phi_{t}+\phi_{PD6:5})   R(\pi,\phi_{t}+\phi_{PD6:6})  R(\pi,\phi_{t}+\phi_{PD6:6})  R(\pi,\phi_{t}+\phi_{PD6:5})  R(\pi,\phi_{t}+\phi_{PD6:4}) \\
R(\pi,\phi_{t}+\phi_{PD6:3}) R(\pi,\phi_{t}+\phi_{PD6:2}) R(\pi,\phi_{t}+\phi_{PD6:1})       
\end{multline}     
with $\phi_{PD6:k}$ given for all $k$'s for $\phi_{t}=\pi$ and $\pi/2$ in Table \ref{tab:pd6}.

The direct impact of Raman laser intensity fluctuation is to modify the amplitude of the qubit rotation, which is calibrated prior to each data set by fitting 3-5 periods of Rabi flopping. Raman laser intensity fluctuations also produce detuning errors due to a differential Stark shift between the two qubit levels. To limit the differential AC Stark shift, the power in each Raman beam is made roughly equivalent on the photodiode (PD in Fig.~\ref{fig:fig1}) by changing the amplitude of $f_1$ and $f_2$ to ensure equal fiber coupling of both Raman beams. The hyperfine frequency, modified by the differential Stark shift, is found using a Ramsey interference experiment.  We begin by initializing the qubit to the $\ket{0}$ state.  The qubit is then placed into a superposition state using a $\theta=\pi/2$ microwave Ramsey pulse. This is followed by a wait time of up to 20~ms, during which Raman beams, 1.5~MHz off-resonant from the qubit frequency, are turned on allowing the differential Stark shift to modify the qubit frequency.  Then the qubit is rotated again with a second $\theta=\pi/2$ microwave Ramsey pulse. The final state of the ion will oscillate with the wait time when the microwave is off-resonant from the qubit frequency. The exact hyperfine qubit frequency in the presence of the Raman beams is found by scanning the microwave frequency to where the ion state is rotated completely to $\ket{1}$ at all wait times. This microwave frequency is used to compute the driving frequency $f_1$ for AOM 2. Although we carefully calibrate the differential Stark shift, drifts in the individual Raman beam amplitudes cause small changes in the qubit frequency ($<100$~Hz), which is not effectively compensated by the pulse sequences used in our experiments.

In a randomized benchmarking experiment, the qubit is first initialized to the $\ket{0}$ state, followed by a sequence of $L$ gate operations, chosen uniformly and randomly from the 24 Clifford gates (shown in Table~\ref{tab:Cl}). A final Clifford gate is then chosen to bring the resulting qubit state to either the $\ket{0}$ or $\ket{1}$ state, at random. Then, the qubit state is measured and compared to the expected state.  For each sequence length $L$, 20 random sequences were created and each sequence was measured 800 times, and the fraction of events where the measured result matched the expected result was recorded as the survival probability.

The averaged survival probability per gate length is fit to the zero-order decay model~\cite{Magesan2012a}
\begin{equation}\label{eq:eq1}
F_{seq}(L)=A_0p^L+B_0,
\end{equation} 
where $F_{seq}(L)$ is the survival probability at length $L$, $p$ is related to the average error per Clifford gate (average error = $\frac{1-p}{2}$), and $A_0$ and $B_0$ contain the state preparation and measurement (SPAM) errors and the error on the final rotation.

\begin{figure}
\begin{center}\includegraphics[width=0.9\textwidth]{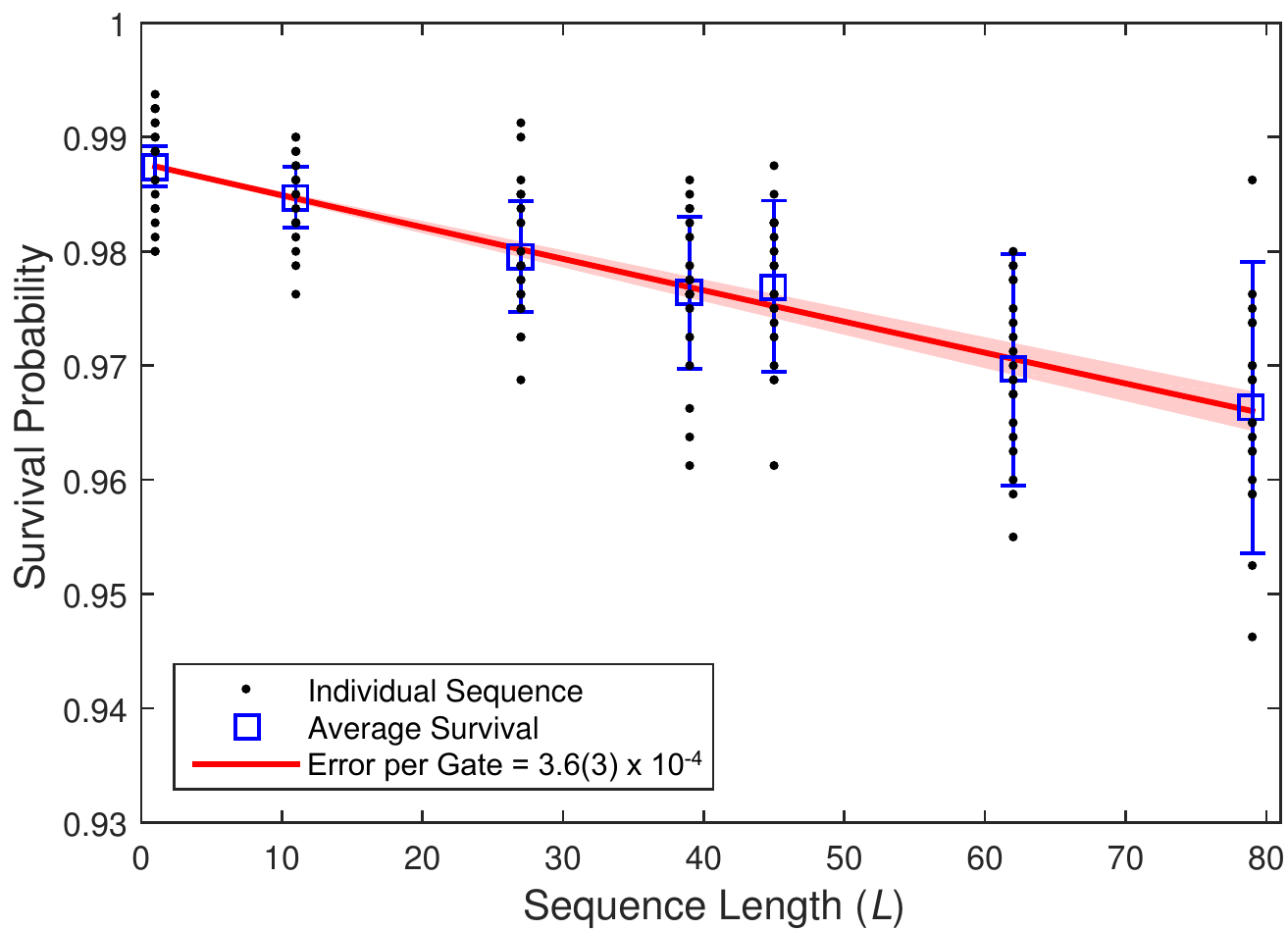}
\caption{
Randomized benchmarking of a single-qubit gate using B2 compensating pulse sequence, with average error per Clifford group gate of $3.6(3) \times 10^{-4}$.
}
\label{fig:fig2}
\end{center}
\end{figure}

The error on our measurement was calculated to account for both the variance due to the projective measurement statistics and the variance arising from the spread in fidelities of the underlying distribution of gate sequences, as outlined in \cite{Wallman2014}. An initial, unweighted, non-linear least squares fit with estimated SPAM errors ($A_0=0.47$ and $B_0=0.517$) was used to gain a first estimate of the underlying error, and from this an upper bound on the variance is calculated. This estimated variance is used to weight a second non-linear least-squares fit with floating SPAM parameters, and the resulting co-variance matrix was used to calculate the uncertainty. Figure~\ref{fig:fig2} shows the result of a randomized benchmarking experiment for a single-qubit gate using the B2 compensating pulse sequence. Black dots represent the average survival probability of each random sequence measured. The average survival probability of all randomized sequences for each sequence length $L$ (blue squares), with an upper-bound variance calculated as in Ref.~\cite{Wallman2014}, are fit to equation~(\ref{eq:eq1}) (red line) to extract an average error per Clifford group gate of $3.6(3)\times10^{-4}$. The error on the fit (light red) accounts for the distribution of measurements and the number of sequences measured for each $L$~\cite{Wallman2014}.

While we expect the actual detuning of the resonant combs arising from the change in the differential AC Stark shift due to amplitude drifts of the Raman beams to be small ($\leq 100$~Hz), a much larger effective detuning error exists. 
The co-propagating Raman beams are circularly polarized, thus, other pairs of comb teeth (\tc{1}-\tc{2} pair,  \tc{3}-\tc{4} pair, and \tc{1}-\tc{4} pair in Fig.~\ref{fig:fig1}b) can also induce Raman transitions, detuned by $\delta' \sim 4.5$~MHz (for the case of \tc{1}-\tc{2} pair, and  \tc{3}-\tc{4} pair) and 9~MHz (for the case of \tc{1}-\tc{4} pair) in our setup. Further off-resonant beat-notes are also present, but have small contribution to gate errors. The contribution of these off-resonant Raman transitions add additional rotation to the desired state evolution, which is calculated by considering  additional terms in the interaction Hamiltonian that describes the qubit subject to the driving field
\begin{eqnarray}\label{eq:hamil}
H_I & = & \frac{\Omega}{2} \{ (X \cos \phi + Y \sin \phi) + 2 [X \cos (\delta' t + \phi_1) + Y \sin (\delta' t + \phi_1)]  \nonumber \\
& + &  [X \cos (2\delta' t + \phi_2)  + Y \sin(2\delta' t + \phi_2)] \},
\end{eqnarray}
where the rotations caused by the detuned Raman transitions due to the \tc{1}-\tc{2} and \tc{3}-\tc{4} pair act coherently on the qubit, and $\phi_1$ and $\phi_2$ describe the relative phases between the detuned Raman transitions and the resonant transition, which fluctuate with drifts in the repetition rate of the laser. The unitary propagator that describes time evolution of the qubit subject to this time-dependent interaction Hamiltonian can be computed using the standard Magnus expansion technique~\cite{Magnus1954}. We use the second-order Magnus expansion to compute the time evolution operator for the qubit subject to each optical pulse. Once the fast oscillating terms (at 4.5MHz and 9MHz) are averaged out, the net impact of the off-resonant terms result in an effective time evolution operator $R(\theta, \phi) =\exp \left (-\frac{i}{2} \theta [\sigma_{\phi} + \delta Z] \right )$, where $\delta Z$ represents a deviation of the qubit rotation axis from the $X-Y$ plane.

For comparison with the experimental data, we simulated the exact benchmarking sequences used in the experiments in the presence of these imperfections. To simulate each sequence of gates, we calculate the imperfect propagator $R_s(\theta, \phi)$ that describes the state evolution of each pulse, where $\theta$ represents the rotation angle and $\phi$ represents the phase of the driving pulse. Increase in pulse area due to static amplitude noise or timing error in the Raman beams leads to an error in the rotation angle of $\theta \to \theta(1 + \epsilon)$, where the multiplicative parameter $\epsilon$ is the same for each pulse in a composite pulse sequence. 
The presence of off-resonant Raman beams and the (smaller) effect of the differential AC Stark shift are captured in the qubit detuning error $\delta$. To simulate a wider range of errors in the pulse area, we purposefully change the timing of the pulses yielding the propagator
\begin{equation}\label{eq:prop}
R(\theta, \phi) =\exp [ -\frac{i}{2} \theta (1 + \epsilon) (\sigma_{\phi} + \delta Z)].
\end{equation}

Our calculation shows that $\delta$ is effectively a random variable in the range of 0-3~kHz due to the drift in the repetition rate of the laser. We calculate a single propagator for each randomized benchmarking sequence by matrix multiplication of the imperfect propagators representing each individual pulse. The resulting final propagators are used to compute the final Bloch state, which is then compared to the expected state yielding a simulated survival probability. The simulated survival probability is fit to equation~(\ref{eq:eq1}) to produce an average error per gate for each series of sequences given specific noise parameters, which is then compared with the experimental results.

\begin{figure}
\begin{center}\includegraphics[width=0.9\textwidth]{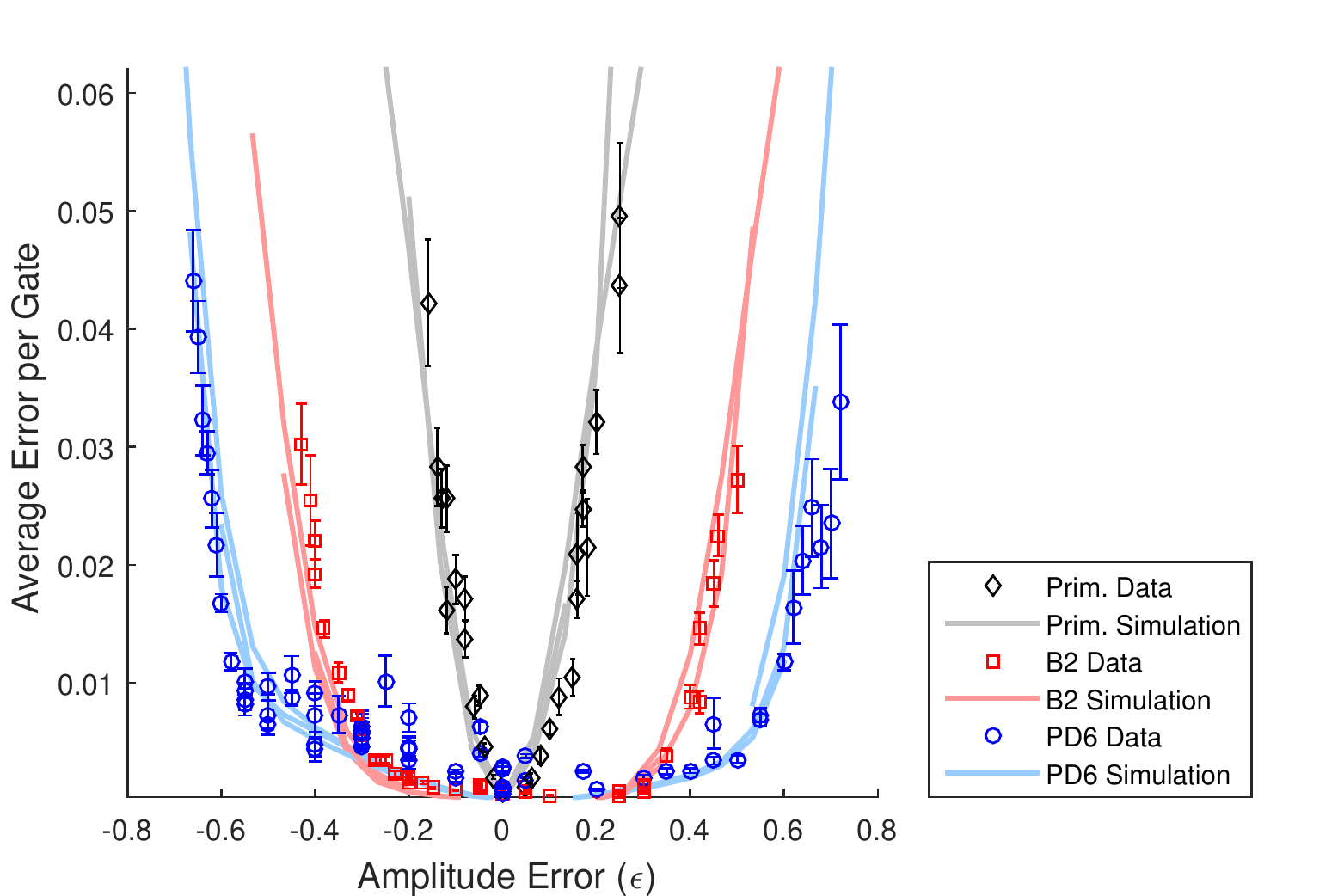}
\caption{
Single-qubit gate error degradation with systematic amplitude error for primitive (black diamonds), B2 compensated (red squares), and PD6 compensated gates (blue circles).  For some amplitude error values, multiple series of randomized sequences were measured resulting in multiple simulation lines over some amplitude error values. 
}
\label{fig:fig3}
\end{center}
\end{figure}

Figure~\ref{fig:fig3} shows a comparison of the error from uncompensated (primitive) gates and gates compensated using B2 and PD6 pulse sequences as a function of timing error. The B2 sequences can keep the gate errors to below 1\% for timing error values in the range of $\mid \epsilon \mid<0.4$, while the PD6 sequences can maintain similar gate error levels for timing error values of up to $\mid \epsilon \mid<0.6$. Our simulations quantitatively reproduce the range over which the timing errors are compensated for both B2 and PD6 sequences using the imperfect propagator computed from the interaction Hamiltonian given in equation~(\ref{eq:hamil}). The minimum error value per gate is achieved by B2 sequence in our setup, since the longer sequence length of PD6 makes it more susceptible to the additional errors contributed by off-resonant Raman transitions.

By driving the gates using two beams with polarizations that are orthogonal to each other and the quantization axis, the intra-comb contributions for the off-resonant Raman transitions (pairs \tc{1}-\tc{2} and \tc{3}-\tc{4}) can be eliminated. The remaining leading-order error will come from inter-comb beat notes (pair \tc{1}-\tc{4}) that can be further detuned by adequate choice of the repetition rate of the mode-locked laser.  Our simulations suggest that these modifications can reduce the average error in the single-qubit gates by over one order of magnitude, at which point the gate performance will be limited by other detuning errors (such as the differential AC Stark shift due to amplitude drifts).
 
In this work, we report high fidelity single-qubit gates driven with tightly focused laser beams on trapped ion qubits by laser intensity stabilization and use of compensating pulse sequences. An error probability as low as $3.6(3)\times10^{-4}$ is demonstrated~\cite{unpub}, consistent with error levels required for realizing a range of quantum error-correction schemes~\cite{Fowler2012,Steane2003,Raussendorf2007}. We experimentally verified the value of novel, length-efficient pulse sequences (PD{\it n}) that suppress errors to higher orders with modest sequence lengths.

The authors would like to acknowledge helpful discussions with True Merrill and Jonathan Mizrahi. This work was supported by the Office of the Director of National Intelligence and Intelligence Advanced Research Projects Activity through the Army Research Office, the ARC via EQuS project number CE11001013, and by the US Army Research Office grant numbers W911NF-14-1-0098 and W911NF-14-1-0103. STF also acknowledges support from an ARC Future Fellowship FT130101744.

\clearpage
\begin{table}
\centering
\resizebox{0.6\textwidth}{!}{
\begin{tabular}{ c |  c  |  c |  c  |  c |  c  |  c}
$\phi_{t}$ & $\phi_{PD6:1}$ & $\phi_{PD6:2}$ & $\phi_{PD6:3}$ & $\phi_{PD6:4}$ & $\phi_{PD6:5}$ & $\phi_{PD6:6}$\\
 \hline
$\pi$ & 0.38266 & -2.51430 & -1.75192 & 0.05941 & 2.67572 & 0.39344 \\
$\pi/2$ & 0.34769 & -3.06979 & 1.55852 & -0.70890 & 3.09692 & -0.62174\\
 \hline
 \end{tabular}
}
\caption{PD6 Phase Angles}
\label{tab:pd6}
\end{table}

\clearpage
\setcounter{table}{0}
\begin{table}[!ht]
\centering
\resizebox{0.6\textwidth}{!}{
\begin{tabular}{ c |  c  |  c |  c }
 Clifford gate &  Physical gates  & Clifford gate &  Physical gates \\
 \hline
1 & I &	13 & Z/2 \& X\\
2 & X&	14 & X \& Z/2\\
3 & Y &	15 & Z/2 \& X/2 \\
4 & Z &	16 & Y/2 \& Z/2 \\
5 & X/2 &	17 & X/2 \& -Z/2 \\
6 & Y/2 &	18 & Y/2 \& Z \\
7 & Z/2 &	19 & -X/2 \& Z/2 \\
8 & -X/2 &	20 & -Z/2 \& Y/2 \\
9 & -Y/2 &	21 & Z \& Y/2 \\
10 & -Z/2 &	22 & -Z/2 \& X/2\\
11 & Z \& X/2 &	23 & X/2 \& Z/2 \\
12 & X/2 \& Z &	24 & -Y/2 \& -Z/2 \\
 \hline
 \end{tabular}
}
\caption{Clifford group gates written as the physical gates (pulses) applied.}
\label{tab:Cl}
\end{table}

\clearpage
\clearpage

\clearpage
\clearpage

\clearpage
\clearpage


\clearpage

\end{document}